\documentclass[letterpaper,journal]{IEEEtran}
\usepackage{amsmath,amsfonts,amssymb}
\usepackage{array}
\usepackage{booktabs}
\usepackage{graphicx}
\usepackage{cite}
\usepackage{stfloats}
\usepackage{needspace}
\usepackage{url}
\usepackage[hidelinks]{hyperref}
\usepackage{float}
\newcommand{\PICEtablefont}{\scriptsize}
\newcommand{\PICEfigurecaptionspacing}{\par\vspace{-1pt}\setlength{\abovecaptionskip}{2pt}}

\begin{document}

\title{PICE: Prototype-Informed Conditional Model Evolution Toward Self-Evolving Medical Image Recognition}

\author{Yifan Zhu, Jie Xu, Zeju Li, Yi Guo,~\IEEEmembership{Member,~IEEE,} and  Yuanyuan Wang, ~\IEEEmembership{Senior Member,~IEEE}
	\thanks{This work was supported by the National Natural Science Foundation of China (Grant 62531004).(Corresponding authors: Yi Guo; Yuanyuan Wang.)}
	\thanks{Yifan Zhu, Jie Xu, Zeju Li, Yi Guo, and Yuanyuan Wang are with the College of Biomedical Engineering, Fudan University, Shanghai 200433, China (email: yifanzhu25@m.fudan.edu.cn; xujie23@m.fudan.edu.cn; zejuli@fudan.edu.cn; guoyi@fudan.edu.cn; yywang@fudan.edu.cn).}}

\maketitle

\begin{abstract}
Medical image classifiers in clinical practice must continually evolve to accommodate emerging disease categories while maintaining reliable classification of existing conditions. This is challenging because privacy and data-sharing restrictions limit access to historical patient data, while assigning dedicated capacity to each new diagnostic task causes linear model growth. Effective evolution requires preserving prior disease knowledge without retaining historical images while expanding model capacity only when necessary. To this end, we propose PICE, a Prototype-Informed Conditional Model Evolution framework organized around what to preserve, when evolution is needed, and how that evolution is realized. PICE preserves prior disease knowledge in a prototype-statistical memory. For each incoming task, it assesses compatibility with historical references reconstructed from memory, reusing the closest adapter when compatibility is sufficient and instantiating a new one when evolution is required. PICE then realizes evolution by updating the memory and recalibrating the unified classifier with Gaussian pseudo features, thereby integrating new knowledge into the cumulative diagnostic space. During inference, memory-derived similarities guide task-agnostic adapter selection and fusion, enabling prediction over all learned disease categories. Experiments on Skin8, Kvasir-V2, and Path16 demonstrate the effectiveness of PICE across diverse clinical scenarios. On Skin8, PICE achieves a 7.60\% improvement in accuracy over the SOTA method on new tasks, while maintaining 71.10\% accuracy on previous tasks. It also reduces additional learnable parameters by 18.51\%. These results indicate that PICE improves classification while balancing knowledge retention, adaptation, and parameter growth, supporting continual model evolution as new diagnostic categories emerge without access to historical images.
\end{abstract}


\section{Introduction}\label{introduction}

\IEEEPARstart{D}{espite} the success of deep learning in medical artificial intelligence (AI), most existing models still rely on a static offline training assumption, where all data are available in a single training stage and model parameters remain unchanged after deployment~\cite{ref1}. However, clinical data streams continually evolve with emerging diseases, changing imaging protocols, and diverse patient populations~\cite{ref2}. Models must learn from these newly collected data, but privacy and data-sharing constraints often prevent access to historical patient images, making it difficult to retain prior diagnostic knowledge and leading to catastrophic forgetting~\cite{ref3}. Assigning dedicated parameters to each new task is a viable solution, but it leads to linear parameter growth and increases the burden of long-term clinical deployment. These challenges motivate a self-evolving paradigm for medical AI, in which models continually reorganize their capacity and update diagnostic knowledge as new categories emerge~\cite{ref4}.

Class incremental learning (CIL) offers a practical formulation for this goal by learning emerging disease categories while preserving established knowledge~\cite{ref36}. Existing CIL methods can be broadly divided into replay-based methods and replay-free methods~\cite{ref5}. Replay-based methods preserve old knowledge by rehearsing stored samples~\cite{ref6,ref7}, but their dependence on historical data conflicts with medical data constraints. Replay-free methods instead adapt to new tasks without revisiting previous images, making them more suitable for privacy-sensitive medical applications. However, updating all model parameters may disrupt learned representations and exacerbate forgetting, while large scale pretrained models (PTMs) provide robust, transferable features by keeping the backbone frozen and adapting incoming tasks through minimal fine-tuning~\cite{ref8}. Among PTMs methods, prompt-based methods typically insert lightweight learnable vectors to adapt a frozen backbone to sequential diagnostic tasks~\cite{ref9,ref10}, yet these prompts may be insufficient to characterize the heterogeneous medical information~\cite{ref11}. Meanwhile, adapter-based methods introduce bottleneck modules into pretrained backbones, providing a more flexible way to adapt diagnostic representations~\cite{ref12,ref13}. Depending on how adaptation capacity is organized, existing adapter-based methods either reuse a shared adapter across all diagnostic tasks~\cite{ref14} or allocate a task-specific adapter to each incoming task~\cite{ref33}. For task-specific designs, the retained adapters are selected or combined during inference through fixed routing, uniform averaging, or learnable routing~\cite{ref16,ref17,ref18}.

Despite recent progress, adapter-based replay-free medical CIL still lacks a controlled evolution mechanism for accommodating emerging disease categories while maintaining accurate recognition of previously learned diseases. This challenge involves three closely related aspects. (1) Lack of historical references. Without previous medical images, the model cannot directly characterize prior disease distributions or assess their relationship with incoming data~\cite{ref19}. (2) Capacity allocation dilemma. A single shared adapter requires few additional parameters but may not capture heterogeneous medical distributions, whereas expanding a new adapter for each task leads to linear parameter growth and redundant modules~\cite{ref20}. (3) Classifier bias after adaptation. Training the unified classifier only on current medical images can shift old--new decision boundaries toward emerging diseases, resulting in serious catastrophic forgetting~\cite{ref21}.

To this end, we propose PICE, a Prototype-Informed Conditional Model Evolution framework organized around what to preserve, when evolution is needed, and how that evolution is realized. Without retaining previous images, PICE preserves prior disease knowledge in a compact prototype--covariance memory to reconstruct historical references for subsequent learning. To address the capacity allocation dilemma, PICE uses reconstructed references to determine whether to reuse an existing adapter or instantiate a new one, enabling controlled model evolution without unnecessary parameter growth. After adaptation, PICE updates the statistical memory and recalibrates the classifier with Gaussian pseudo features to mitigate classifier bias and consolidate both new and prior disease knowledge. During inference, memory-guided adapter selection and fusion enable task-agnostic prediction over all learned disease categories. In this work, our main contributions can be summarized as follows:

\begin{itemize}
\item We propose PICE, a replay-free medical CIL framework that supports controlled self-evolving for emerging disease recognition while maintaining reliable performance on previously learned diseases.

\item We develop a prototype-informed evolution mechanism that preserves what is essential from prior disease distributions in prototype--covariance memory, identifies when structural evolution is needed through task compatibility assessment, and determines how the model evolves by conditionally allocating capacity, memory updating and classifier recalibration with Gaussian pseudo features.

\item Extensive experiments across diverse medical benchmarks demonstrate that PICE achieves superior performance on newly introduced tasks while effectively preserving performance on prior tasks, and evolves with substantially lower parameter growth compared to state-of-the-art CIL methods.

\end{itemize}

\section{\texorpdfstring{Related work }{Related work }}\label{related-work}

\subsection{Replay-Based Medical CIL}\label{replay-based-medical-cil}

Significant efforts have been made to develop CIL techniques for natural and medical image analysis~\cite{ref5}. However, directly applying CIL to medical imaging remains challenging because clinical data streams are often affected by evolving acquisition protocols, scanner variations, institutional heterogeneity, and the continual emergence of new disease categories~\cite{ref3}. Existing CIL methods usually rely on replay-based or replay-free strategies. Replay-based methods alleviate forgetting by storing a subset of previously seen samples and jointly training them with current task data~\cite{ref6,ref7}, which is effective but often impractical in clinical applications due to privacy, data-sharing, and storage constraints~\cite{ref22}. These limitations make replay-free CIL a more suitable direction for medical AI, as it seeks to preserve previously acquired diagnostic knowledge without storing historical patient images.

\subsection{Prompt-Based Replay-Free PTM CIL}\label{prompt-based-replay-free-ptm-cil}

Conventional replay-free methods often update the full model using current task images, with regularization or knowledge distillation to preserve prior knowledge. Regularization penalizes changes to parameters important for previous tasks, whereas knowledge distillation constrains current predictions or features to remain consistent with the previous model~\cite{ref5,ref34}. However, without historical images, these full model parameters update may seriously overwrite disease-specific representations. With the increasing availability of PTMs, recent CIL methods have adopted architectures such as ViT and CLIP as backbones to leverage transferable representations for sequential task adaptation~\cite{ref8,ref14}. Building on these pretrained backbones, parameter-efficient tuning adapts PTMs to sequential medical imaging tasks through lightweight components such as prompts and adapters. Prompt-based methods insert a small set of learnable vectors into a frozen PTM to guide feature extraction for different diagnostic tasks~\cite{ref9,ref10}. Each prompt is paired with a key, and the feature of an input image serves as a query to retrieve the prompts with the most similar keys during inference. However, these lightweight vectors may not fully capture subtle and heterogeneous disease manifestations. Moreover, key-query matching can retrieve inappropriate prompts when visual patterns overlap across disease categories or vary with acquisition protocols, scanner conditions, and institutional characteristics, thereby compromising downstream predictions~\cite{ref3,ref37}.

\subsection{Adapter-Based Replay-Free PTM CIL}\label{adapter-based-replay-free-ptm-cil}

Unlike prompts that guide a frozen backbone through additional learnable tokens, adapters insert small trainable bottleneck layers within the network to directly transform intermediate diagnostic features~\cite{ref12,ref13}. Their greater representational flexibility makes the allocation and management of adapters particularly important in sequential medical imaging. Shared-adapter methods reuse a common adapter across tasks to promote general knowledge transfer, but they may suffer from inter-task interference under heterogeneous medical data streams~\cite{ref14,ref31}. Task-specific adapters allocate dedicated modules for newly arriving tasks to preserve specialized knowledge and mitigate forgetting, but they typically incur growing parameter costs and limited knowledge transfer across tasks~\cite{ref18}. Neither strategy determines whether an incoming distribution can be adequately represented by existing capacity before expanding the model. When multiple adapters are retained, inference must also determine how to use them without task identity. Some methods select a single relevant adapter for each input, but such hard selection can be unreliable when task boundaries are ambiguous~\cite{ref16}. Alternatively, adapter fusion methods combine multiple adapters to exploit complementary task knowledge, yet fixed weighting, simple averaging, or trainable routers may still fail to capture sample-specific relationships between incoming representations and accumulated diagnostic knowledge when neither task identity nor historical medical images are available~\cite{ref23,ref24}. Overall, existing adapter-based methods rely on predefined sharing, expansion, or fusion strategies, lacking a conditional mechanism to evolve model capacity according to incoming disease distributions and previously acquired knowledge.

\section{Methodology}\label{methodology}

\begin{figure*}[!t]
	\centering
	\includegraphics[width=\textwidth,trim=0 14bp 0 0,clip]{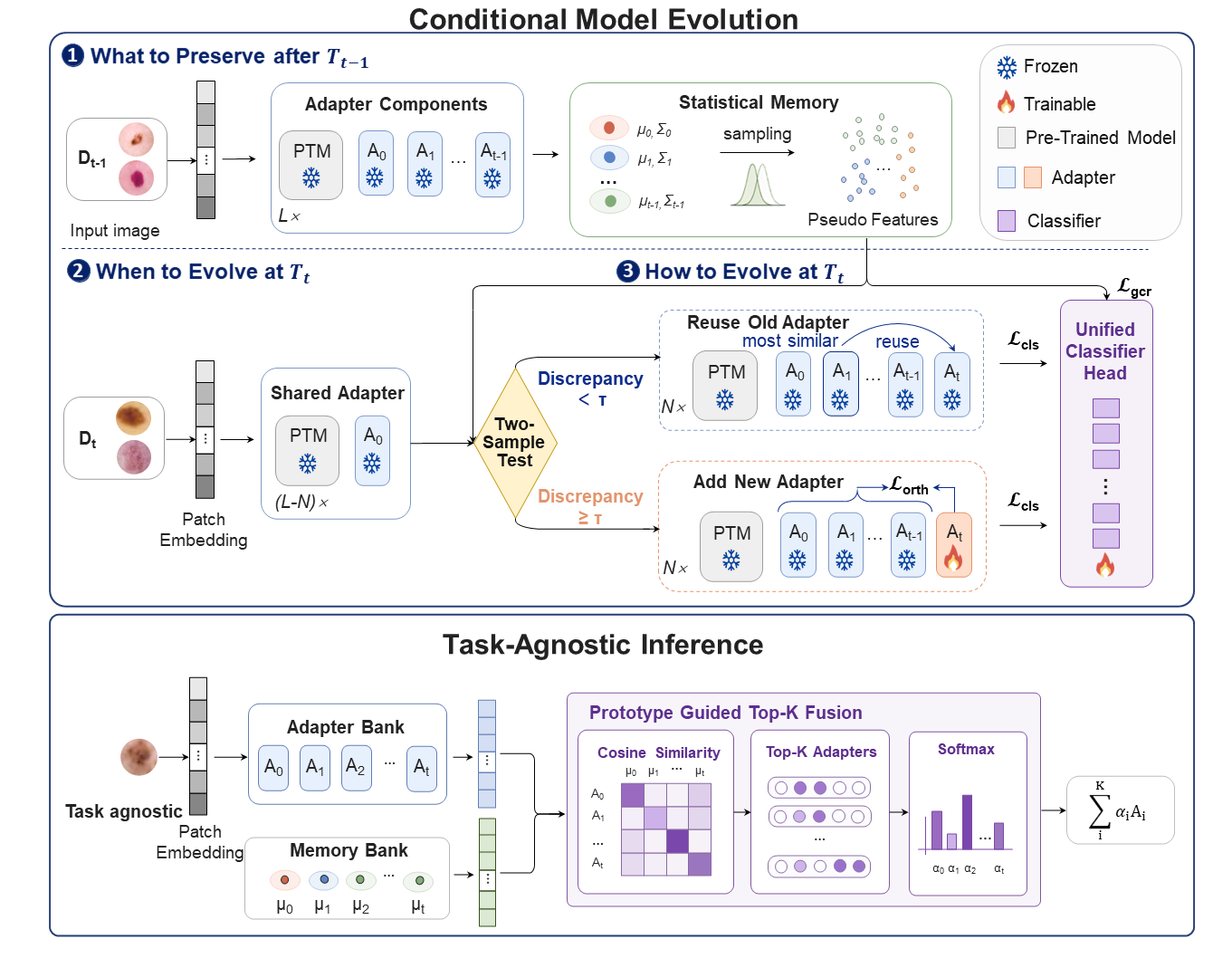}
	\PICEfigurecaptionspacing
	\caption{Overall framework of PICE. After learning \(T_{t-1}\), prior knowledge is retained as prototype-statistical memory. For \(T_t\), C2ST compares current features with historical references reconstructed from memory to guide deep adapter allocation, while shallow adapters remain shared. After adaptation, memory updating and Gaussian classifier recalibration consolidate knowledge. During task-agnostic inference, prototype similarity guides Top-K adapter fusion.}
	\label{fig:1}
\end{figure*}

\subsection{Problem Formulation and Framework Overview}\label{problem-formulation-and-framework-overview}

Class incremental learning aims to enable a model to continually evolve as new disease categories are introduced, while preserving recognition of previously learned categories. In practice, \(\{\mathcal{D}_{t}\}_{t = 1}^{T}\) denotes a sequence of training datasets acquired in temporal order, where \(\mathcal{D}_{t} = \{(x_{i}^{t},y_{i}^{t})\}_{i = 1}^{N_{t}}\) contains \(N_{t}\) image-label pairs for the current task \(T_{t}\). When a new task \(T_{t}\) is introduced, only \(\mathcal{D}_{t}\) is accessible, while all previously observed datasets are unavailable and not retained. Each label \(y_{i}^{t}\) belongs to the class set \(\mathcal{C}_{t}\), where \(\mathcal{C}_{t}\) denotes the new disease categories introduced at \(T_{t}\), with \(\mathcal{C}_{t} \cap \mathcal{C}_{k} = \varnothing\) for \(k < t\). After learning the new task \(T_{t}\), the model is expected to classify samples over the accumulated label space \(\mathcal{Y}_{t} = \bigcup_{k = 1}^{t}\mathcal{C}_{k}\), without access to the task identity.

Here, before continual evolution begins, the model is initialized on the old task using an ImageNet-pretrained Vision Transformer with \(L\) Transformer blocks. The backbone remains frozen, while lightweight bottleneck adapters are inserted into the Transformer blocks to provide trainable capacity~\cite{ref25}. Given the input feature \(x_{\text{in}}^{l}\) of the \(l^{th}\) block, the output after adaptation is computed as

\begin{equation}
x_{\mathrm{out}}^{l}=x_{\mathrm{in}}^{l}+\operatorname{MLP}(x_{\mathrm{in}}^{l})
+\gamma\,\operatorname{ReLU}(x_{\mathrm{in}}^{l}W_{\mathrm{down}}^{l})W_{\mathrm{up}}^{l}.
\label{eq:1}
\end{equation}

where \(W_{\text{down}}^{l} \in \mathbb{R}^{d \times r}\) and \(W_{\text{up}}^{l} \in \mathbb{R}^{r \times d}\) are the down projection and up projection matrices, \(d\) is the hidden dimension, \(r \ll d\) is the bottleneck dimension, and \(\gamma\) controls the adapter contribution. For the initial task \(T_{1}\), the adapters and unified classifier are optimized on \(\mathcal{D}_{1}\) while the pretrained backbone remains frozen, adapting the ImageNet-pretrained representation to the medical image domain and establishing the initial model for subsequent evolution.

Building on this initialization, we propose PICE, a Prototype-Informed Conditional Model Evolution framework for replay-free medical CIL, organized around what to preserve, when to evolve, and how to evolve. Fig.~\ref{fig:1} summarizes the resulting evolution process. PICE first preserves prior disease distributions in prototype-statistical memory, providing historical references when old task images are unavailable. When a new task \(T_{t}\) is introduced, a classifier two-sample test (C2ST) compares its distribution with historical references to determine when adapter capacity should evolve through reuse or expansion. The allocated capacity is then adapted to \(T_{t}\), followed by memory updating and classifier recalibration for knowledge consolidation to realize model evolution. During inference, memory-guided adapter selection and fusion enable task-agnostic prediction over all learned categories.

\subsection{\texorpdfstring{What to Preserve: Prototype-Statistical Memory }{What to Preserve: Prototype-Statistical Memory }}\label{what-to-preserve-prototype-statistical-memory}

After learning each task with the frozen backbone and its assigned adapters, the corresponding training images become unavailable under the replay-free setting. PICE therefore maintains compact feature statistics for each learned class rather than retaining raw samples. Specifically, for each class \(c \in \mathcal{C}_{t}\) introduced in \(T_{t}\), we store a prototype \(\mu_{t,c}\) and a covariance matrix \(\Sigma_{t,c}\), which characterize the center and intra class variation of its feature distribution~\cite{ref26}. Let \(F_{\theta,\mathcal{A}_{t}}( \cdot )\) denote the feature extractor composed of the frozen backbone and the adapter collection \(\mathcal{A}_{t}\ \)assigned to \(T_{t}\). Since task adapters may induce different feature spaces, class statistics are estimated with the corresponding adapter.

\begin{equation}
\mu_{t,c}=\frac{1}{N_{t,c}}\sum_{i:y_i^t=c}F_{\theta,\mathcal{A}_t}(x_i^t).
\label{eq:2}
\end{equation}

where \(N_{t,c}\) denotes the number of training samples belonging to class \(c\) in \(\mathcal{C}_{t}\). The covariance matrix is estimated in the same adapter-induced feature space as

\begin{equation}
\begin{aligned}
\Sigma_{t,c}={}&\frac{1}{N_{t,c}-1}\sum_{i:y_i^t=c}
\bigl(F_{\theta,\mathcal{A}_t}(x_i^t)-\mu_{t,c}\bigr)\\
&\times\bigl(F_{\theta,\mathcal{A}_t}(x_i^t)-\mu_{t,c}\bigr)^{\top}+\lambda I.
\end{aligned}
\label{eq:3}
\end{equation}

where \(\lambda\) is a regularization coefficient for numerical stability. After learning \(T_{t}\), the accumulated prototype-statistical memory is defined as
\begin{equation}
M_t=\{(\mu_{k,c},\Sigma_{k,c})\mid c\in\mathcal{C}_k,\ k\le t\}.
\label{eq:4}
\end{equation}

This prototype-statistical memory adaptively determines what to preserve during self-evolving, by providing a compact representation of previously learned disease distributions, without retaining historical patient images. It further serves as the preserved knowledge base for reconstructing historical references to determine when structural evolution is needed and for sampling Gaussian pseudo features during subsequent knowledge consolidation.

\subsection{When to Evolve: Conditional Adapter Allocation}\label{when-to-evolve-conditional-adapter-allocation}

Having defined the historical knowledge preserved in memory, PICE then determines when additional capacity is required for a new task. To this end, it first restricts the candidate capacity for structural evolution via a layer-aware adapter architecture, then evaluates compatibility between the new task and historical references to decide whether existing adapter capacity suffices or structural expansion is needed.

\subsubsection{\texorpdfstring{\textbf{Layer-Aware Adapter Architecture}}{Layer-Aware Adapter Architecture}}\label{layer-aware-adapter-architecture}

To prevent unrestricted structural expansion, PICE employs a layer-aware adapter architecture that confines structural evolution to deep Transformer blocks while sharing shallow adapters across tasks. For the shallow part, we insert shared adapters into the first \(L_{s}\) Transformer layers and denote their collection as \(\mathcal{A}^{\text{shared}} = \{ A_{l}^{\text{shared}}\}_{l = 1}^{L_{s}}\). In the remaining blocks, the deep adapters serve as task-specific expert capacity and are denoted by \(\mathcal{A}_{t}^{\text{expert}} = \{ A_{t,l}^{\text{expert}}\}_{l = L_{s} + 1}^{L}\). The complete adapter collection associated with \(T_{t}\) is then written as \(\mathcal{A}_{t} = \mathcal{A}^{\text{shared}} \cup \mathcal{A}_{t}^{\text{expert}}\). PICE maintains this shared--expert organization throughout each learning stage. For the initial task \(T_{1}\), both shared and expert adapters are optimized to establish the initial representation. For subsequent new tasks, \(\mathcal{A}^{\text{shared}}\) is kept frozen and reused, and all subsequent structural evolution is confined to the deep expert adapters, which are conditionally reused or expanded as described below.

\subsubsection{\texorpdfstring{\textbf{C2ST-Based Reuse-or-Expansion Decision}}{C2ST-Based Reuse-or-Expansion Decision}}\label{c2st-based-reuse-or-expansion-decision}

Since historical images are unavailable, we construct reference features from the prototype-statistical memory \(M_{t - 1}\). For each previous task \(k < t\), we sample Gaussian features from \(\mathcal{N}\left( \mu_{k,c},\Sigma_{k,c} \right)\), whose mean and covariance were computed from class\(\text{\ c}\) features extracted by \(\mathcal{A}_{k}\), and extract current task features using the same adapter collection:

\begin{equation}
\left\{
\begin{aligned}
\widetilde{z}_{j}^{k,c}&\sim\mathcal{N}(\mu_{k,c},\Sigma_{k,c}), &&c\in\mathcal{C}_k,\\
z_i^{t\rightarrow k}&=F_{\theta,\mathcal{A}_k}(x_i^t), &&i=1,\ldots,N_t.
\end{aligned}
\right.
\label{eq:5}
\end{equation}

Here, \({\widetilde{z}}_{j}^{k,c}\) denotes the \(j^{\text{th}}\) pseudo feature for class \(c\) in historical task \(k\), while \(z_{i}^{t \rightarrow k}\) denotes the feature of the \(i^{\text{th}}\) incoming image extracted using the historical adapter collection \(\mathcal{A}_{k}\). This construction enables a consistent comparison between the incoming task and historical reference distribution.

PICE then applies a C2ST between current task features
\(\{z_i^{t\rightarrow k}\}_{i=1}^{N_t}\)
and the Gaussian reference features
\(\{{\widetilde{z}}_{j}^{k,c} \mid c \in \mathcal{C}_{k}\}\). The resulting discrepancy score is denoted as \(\Delta_{t,k}\), which measures the distributional separability between the current task and the reference features generated from task \(T_{k}\). A lower score indicates a closer match. We therefore select the most compatible previous task by \(k^{\star} = \text{arg}\text{min}_{k < t}\Delta_{t,k}\). Based on this selected task, the expert adapter collection for the current task is determined by

\begin{equation}
\mathcal{A}_t^{\mathrm{expert}}=
\begin{cases}
\mathcal{A}_{k^{\star}}^{\mathrm{expert}}, & \Delta_{t,k^{\star}}<\tau,\\
\mathcal{A}_{\mathrm{New}}^{\mathrm{expert}}, & \Delta_{t,k^{\star}}\ge\tau.
\end{cases}
\label{eq:6}
\end{equation}

where \(\tau\) is a predefined discrepancy threshold. When \(\Delta_{t,k^{\star}}<\tau\), PICE reuses the expert adapter collection \(\mathcal{A}_{k^{\star}}^{\text{expert}}\) for \(T_{t}\). Otherwise, PICE initializes and optimizes a new expert adapter for \(T_{t}\) using \(D_{t}\).

With statistical memory guiding the allocation decision, PICE determines when structural evolution is required and avoids assigning a new adapter to every task by a fixed rule. This selective evolution limits parameter growth while promoting knowledge reuse across related disease categories.

\subsection{How to Evolve: Adaptation and Knowledge Consolidation}\label{how-to-evolve-task-adaptation-and-knowledge-consolidation}

Once the reuse or expansion decision is made, PICE realizes model evolution through task adaptation followed by knowledge consolidation. Given the adapter collection \(\mathcal{A}_{t}\) assigned to \(T_{t}\), the optimization strategy depends on the selected evolution path. When existing adapter capacity is reused, all adapters remain frozen and only the unified classifier is updated. When structural evolution is triggered, the newly instantiated deep expert adapters are optimized together with the classifier, while other components remain frozen. In implementation, both feature vectors and classifier weights are \(\ell_2\)-normalized before loss
computation, so their inner products represent cosine similarities. The classification objective is: 

\begin{equation}
\begin{aligned}
\mathcal{L}_{\mathrm{cls}}^t={}&-\frac{1}{N_t}
\sum_{(x_i^t,y_i^t)\in\mathcal{D}_t}
\log\frac{\exp\!\left(w_{y_i^t}^{\top}F_{\theta,\mathcal{A}_t}(x_i^t)\right)}
{\sum_{c\in\mathcal{Y}_t}\exp\!\left(w_c^{\top}F_{\theta,\mathcal{A}_t}(x_i^t)\right)}.
\end{aligned}
\label{eq:7}
\end{equation}

where \(w_{c}\) is the classifier weight for class \(c\), and \(\mathcal{Y}_{t} = \bigcup_{k = 1}^{t}\mathcal{C}_{k}\) denotes the accumulated label space.

Once structural evolution is triggered, an orthogonality constraint is imposed between its up-projection weights and those of the existing adapters. This constraint encourages the newly allocated capacity to capture complementary disease-specific imaging characteristics rather than duplicating previously learned representations:

\begin{equation}
\mathcal{L}_{\mathrm{orth}}^t=\sum_{k<t}\left\|W_{\mathrm{up}}^t
\left(W_{\mathrm{up}}^k\right)^{\top}\right\|_1.
\label{eq:8}
\end{equation}

where \(W_{\text{up}}^{t}\) denotes the concatenated up-projection weights of the expert adapters in \(\mathcal{A}_{t}^{\text{expert}}\). The overall objective is:

\begin{equation}
\mathcal{L}^t=\mathcal{L}_{\mathrm{cls}}^t
+\lambda_{\mathrm{orth}}\mathcal{L}_{\mathrm{orth}}^t.
\label{eq:9}
\end{equation}

For the reuse case, no additional adapter is optimized and the orthogonal loss is not applied.

After task adaptation, PICE integrates the newly acquired disease knowledge into its cumulative diagnostic space. The class prototypes and covariance matrices of the newly introduced categories are estimated using the assigned adapter collection according to Eqs.~\eqref{eq:2} and~\eqref{eq:3}, and are incorporated into the prototype-statistical memory \(M_{t}\). During sequential learning, optimizing the unified classifier primarily with current task data can distort decision boundaries across learning stages, causing predictions to become biased toward newly introduced disease categories. To mitigate this bias, PICE performs rehearsal-free Gaussian Classifier Recalibration (GCR) after each task. GCR reuses the sampling operation in Eq.~\eqref{eq:5} to construct a balanced pseudo feature set \({\widetilde{\mathcal{Z}}}_{1:t}^{\text{gcr}}\) for classifier recalibration. During this stage, all feature extractors are frozen and only the classifier is updated by minimizing

\begin{equation}
\begin{aligned}
\mathcal{L}_{\mathrm{gcr}}^t={}&-
\frac{1}{|\widetilde{\mathcal{Z}}_{1:t}^{\mathrm{gcr}}|}
\sum_{(\widetilde{z},y)\in\widetilde{\mathcal{Z}}_{1:t}^{\mathrm{gcr}}}
\log\frac{\exp(w_y^{\top}\widetilde{z})}
{\sum_{c\in\mathcal{Y}_t}\exp(w_c^{\top}\widetilde{z})}.
\end{aligned}
\label{eq:10}
\end{equation}

where \((\widetilde{z},y)\) is sampled from the Gaussian distribution associated with class \(y\). Overall, PICE first adapts the required model capacity to the new task. The acquired disease statistics are then incorporated into the prototype-statistical memory, after which the unified classifier is recalibrated over all learned categories. This coordinated process progressively updates model capacity, diagnostic knowledge, and decision boundaries without retaining historical medical images.

\subsection{\texorpdfstring{Task-Agnostic Inference: Prototype-Guided Top-\(K\) Fusion}{Task-Agnostic Inference: Prototype-Guided Top-K Fusion}}\label{task-agnostic-inference-prototype-guided-top-mathbfk-fusion}

Following the preceding formulation of what to preserve, when to evolve, and how to evolve, PICE performs task-agnostic inference through prototype-guided Top-\emph{K} adapter fusion. During inference, routing each image to a single adapter can be unreliable when its imaging characteristics overlap with several learned disease categories. PICE therefore performs prototype-guided Top-\(K\) adapter fusion, using stored memory to select and combine relevant adapters without an additional router or task identity. Given a test sample \(x\), the relevance of the \(k^{\text{th}}\) adapter collection is measured in its corresponding feature space as

\begin{equation}
s_k(x)=\max_{c\in\mathcal{C}_k}
\operatorname{cos}\!\left(F_{\theta,\mathcal{A}_k}(x),\mu_{k,c}\right),
\quad k=1,\ldots,t.
\label{eq:11}
\end{equation}

PICE retains the \emph{K} adapters with the highest scores and normalizes their scores using softmax to obtain fusion weights \(\alpha_{k}(x)\). Unselected adapters receive zero weight. At each deep Transformer layer, the outputs are fused as

\begin{equation}
	x_{\mathrm{out}}=x_{\mathrm{in}}+\mathrm{MLP}(x_{\mathrm{in}})+\sum_{k=1}^{t}\alpha_k(x)\mathcal{A}_k^{\mathrm{expert}}(x_{\mathrm{in}}).
\label{eq:12}
\end{equation}

The fused representation is then passed to the recalibrated unified classifier, which produces prediction scores over the accumulated label space \(\mathcal{Y}_{t}\). The category with the highest score is returned as the final task-agnostic prediction. This prototype-guided strategy allows relevant diagnostic adapters to contribute jointly while avoiding direct comparisons across inconsistent feature spaces.

\section{Experiments}\label{experiments-1}

\subsection{Datasets}\label{datasets}

\begin{figure}[!t]
\centering
\includegraphics[width=\columnwidth,trim=0 6bp 0 0,clip]{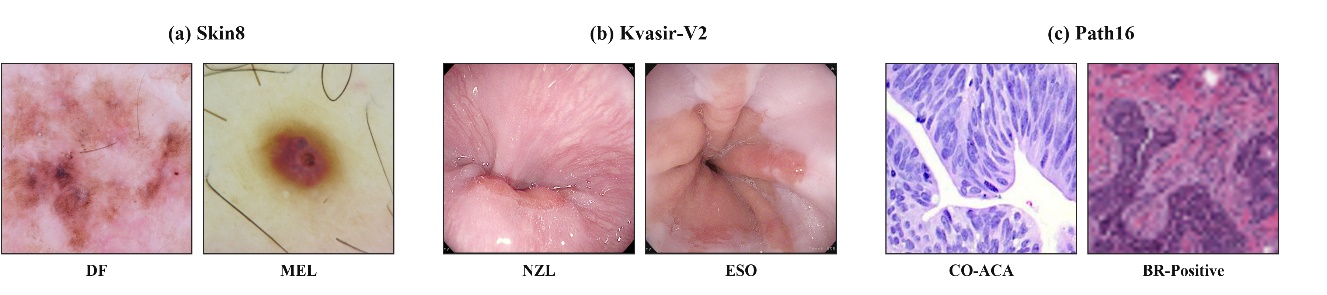}
\PICEfigurecaptionspacing
\caption{Representative images from Skin8, Kvasir-V2, and Path16, with the corresponding categories labeled below each image.}
\label{fig:2}
\end{figure}

We evaluate our method on three medical datasets, including Skin8, Path16, and Kvasir-V2, which span three distinct imaging modalities (dermatology, pathology, and endoscopy) and address diverse clinical applications. Representative images from the three datasets are shown in Fig.~\ref{fig:2}, while their dataset statistics and incremental settings are summarized in Table~\ref{tab:1}.

\textbf{Skin8.} Skin8 is constructed from public dermoscopic images in the ISIC challenge and contains 8 skin lesion categories with a highly imbalanced class distribution~\cite{ref27}, providing a challenging setting for retaining previously learned minority classes. The dataset contains 3,555 training images and 705 test images. The initial task \(T_{1}\) contains dermatofibroma (DF) and melanocytic nevus (NV). Three incremental tasks are subsequently introduced, with \(T_{2}\) comprising benign keratosis-like lesions (BKL) and squamous cell carcinoma (SCC), \(T_{3}\) comprising basal cell carcinoma (BCC) and actinic keratosis (AK), and \(T_{4}\) comprising vascular lesions (VASC) and melanoma (MEL). The model starts with two disease categories and acquires two new categories at each subsequent stage.

\textbf{Kvasir-V2.} Kvasir-V2 contains gastrointestinal endoscopy images acquired during routine clinical practice for multi-class disease recognition~\cite{ref28}, serving as a challenging setting for incremental recognition among visually similar gastrointestinal categories. It consists of 6,400 training images and 1,600 test images. The initial task \(T_{1}\) contains normal cecum (NC) and ulcerative colitis (UC). The subsequent new tasks introduce dyed-lifted polyps (DLP) and dyed resection margins (DRM) in \(T_{2}\), normal z-line (NZL) and esophagitis (ESO) in \(T_{3}\), and normal pylorus (NP) and polyps (POL) in \(T_{4}\). Similar to Skin8, each new task adds two previously unseen categories.

\textbf{Path16.} Path16 is a composite histopathology benchmark introduced in~\cite{ref12}, comprising 16 categories from multiple public datasets across seven tissue-specific incremental tasks, and is used to evaluate continual adaptation under substantial inter-task heterogeneity. The dataset includes 12,808 training images and 1,607 test images. Colon (CO) serves as the initial task with two classes, followed sequentially by six new tasks from lung (LU, 2 classes), lymph node (LY, 2 classes), stomach (ST, 4 classes), oral cavity (OR, 2 classes), colorectal polyp (CP, 2 classes), and breast (BR, 2 classes). Path16 introduces disease categories across distinct tissue domains, resulting in greater heterogeneity between successive tasks.

At incremental stage \(T_{t}\), only the training images associated with the current task are accessible, while images from previously learned tasks are not retained. After each stage, the model is evaluated on all classes observed so far using a single unified classifier. The task identity is unavailable during inference, requiring all methods to perform task-agnostic prediction. For all datasets, input images are processed at a spatial resolution of 224$\times$224 pixels, with random resized cropping and random horizontal flipping applied during training.

\begin{table}[!t]
	\caption{Summary of the three medical image datasets, including imaging modality, number of classes and incremental tasks, and training and test image counts.}
	\label{tab:1}
	\centering
	\PICEtablefont
	\renewcommand{\arraystretch}{1.12}
	\setlength{\tabcolsep}{1.0pt}
	\begin{tabular}{llccll}
		\toprule
		Dataset & Modality & Classes & Tasks & Task-wise Class Split & Train/Test \\
		\midrule
		Skin8 & Dermoscopy & 8 & 4 & 2 / 2 / 2 / 2 & 3,555/705 \\
		Kvasir-V2 & Endoscopy & 8 & 4 & 2 / 2 / 2 / 2 & 6,400/1,600 \\
		Path16 & Pathology & 16 & 7 & 2 / 2 / 2 / 4 / 2 / 2 / 2 & 12,808/1,607 \\
		\bottomrule
	\end{tabular}
\end{table}

\subsection{Compared Methods}\label{compared-methods}

We compare PICE with representative replay-based, prompt-based, and adapter-based class-incremental learning methods, with particular emphasis on adapter-based approaches that are most closely related to our framework.

Among them, CLS-ER~\cite{ref6} and PCR~\cite{ref7} are typical replay-based methods. CLS-ER maintains complementary learning systems through exemplar replay, and PCR combines stored exemplars with proxy-based contrastive learning to preserve class discriminative representations across incremental stages.

For prompt-based PTMs methods, L2P~\cite{ref9} and DualPrompt~\cite{ref10} adapt PTMs with learnable lightweight prompts, where L2P selects prompts from a learned prompt pool, while DualPrompt employs complementary general and expert prompts to capture task-shared and task-specific knowledge.

We further compare five representative adapter-based methods for their relevance to PICE. Adapter-based methods insert trainable bottleneck modules into transformer layers for task adaptation. Specifically, APER~\cite{ref14} adopts a single adapter design. EASE~\cite{ref13} allocates a separate adapter to each task and concatenates the features extracted from all task-specific adapters for joint classification. CRCL~\cite{ref15} employs complementary branches to balance previous knowledge retention and new class learning, whereas TUNA~\cite{ref16} combines task-specific adapters with a universal adapter during inference. ACMap~\cite{ref17} merges task-specific adapters into a shared adapter for scalable incremental learning. These methods represent recent state-of-the-art approaches across replay-based, prompt-based, and adapter-based CIL.

\subsection{\texorpdfstring{Evaluation Metrics }{Evaluation Metrics }}\label{evaluation-metrics}

To comprehensively assess continual model evolution, we evaluate PICE from three complementary perspectives. (1) Classification performance reflects the model's ability to recognize accumulated disease categories throughout model evolution. (2) Stability and plasticity characterize its ability to retain previously learned knowledge while adapting to newly introduced categories. (3) Parameter growth measures the structural cost of accommodating new disease categories during model evolution.

\subsubsection{\texorpdfstring{\textbf{Classification performance} }{Classification performance }}\label{classification-performance}

Let \(a_{k,j}\) denote the accuracy on the test set of \(T_{j}\ \) after learning \(T_{k}\ \), where \(1 \leq j \leq k \leq T\). We use the average accuracy after each incremental stage, the average accuracy throughout the incremental process, and the final accuracy after all tasks have been learned, defined respectively as

\begin{equation}
	\bar{a}_k=\frac{1}{k}\sum_{j=1}^{k}a_{k,j},\quad
	A_{\mathrm{avg}}=\frac{1}{T}\sum_{k=1}^{T}\bar{a}_k,\quad
	A_{\mathrm{last}}=\bar{a}_T.
	\label{eq:13}
\end{equation}

Here, \(A_{\text{avg}}\) reflects classification performance throughout the incremental process, whereas \(A_{\text{last}}\) measures the final classification performance over all learned disease categories.

\subsubsection{\texorpdfstring{\textbf{Stability and plasticity} }{Stability and plasticity }}\label{stability-and-plasticity}

We adopt forgetting and intransigence from~\cite{ref29} to characterize the stability-plasticity trade-off. Specifically, forgetting \(F_{k}\) measures the degradation in performance on old tasks after learning task \(T_{k}\), while intransigence \(I_{k}\) measures the difficulty of adapting to the newly introduced task:

\begin{equation}
\begin{aligned}
F_k&=\frac{1}{k-1}\sum_{j=1}^{k-1}
\left(\max_{l\in\{j,\ldots,k-1\}}a_{l,j}-a_{k,j}\right),\\
I_k&=a_k^{\star}-a_{k,k}.
\end{aligned}
\label{eq:14}
\end{equation}

Here, \(a_{k}^{*}\) denotes the accuracy obtained by a reference model trained jointly on all classes available up to \(T_{k}\). The overall forgetting and intransigence scores, denoted by \(F\) and \(I\), are obtained by averaging \(F_{k}\) and \(I_{k}\) across the applicable incremental stages. Lower \(F\) and \(I\) indicate better stability (retention of previously learned knowledge) and plasticity (adaptation to newly introduced classes). We further summarize their trade-off as

\begin{equation}
S_{\mathrm{FI}}=\sqrt{F^2+I^2}.
\label{eq:15}
\end{equation}

\(S_{\text{FI}}\) measures the Euclidean distance between each method and the ideal point \((0,\ 0)\), with a lower value indicating a better overall balance between stability and plasticity.

\subsubsection{\texorpdfstring{\textbf{Parameter growth} }{Parameter growth }}\label{metric-parameter-growth}

We report the total number of additional trainable parameters beyond the common pretrained backbone after completing all incremental tasks. The parameter count includes the unified classifier for all methods. For adapter-based methods, all shared and task-specific adapters are counted. For prompt-based methods, all learnable prompt parameters are counted. For replay-based methods, only the classifier parameters are counted, with exemplar storage excluded. This metric quantifies the additional structural cost required to accommodate newly introduced disease categories, with fewer added parameters indicating more parameter-efficient model evolution.

\subsection{Training Details}\label{training-details}

We use a ViT-B/16 backbone initialized with ImageNet-21K pretrained weights. The backbone remains frozen throughout incremental learning. During the first task, the adapters are optimized to adapt the pretrained representation to the medical image domain. For subsequent tasks, the unified classifier is updated in both reuse and expansion cases, whereas newly instantiated deep expert adapters are additionally optimized only when structural expansion is triggered. We train the model using stochastic gradient descent with a momentum of 0.9, a weight decay of \(5 \times 10^{- 4}\), and an initial learning rate of 0.05 scheduled by cosine annealing. Adapter expansion is applied only to the last four Transformer blocks, while the remaining adapters are shared across tasks.

For a fair comparison, all methods are trained and evaluated under the same experimental protocol. The exemplar budgets of CLS-ER and PCR are fixed at 500. For L2P, the prompt length and pool size are set to 10 and 16. For DualPrompt, the expert prompt length and pool size are set to 10 and 16, while the shared general prompt length is set to 5. For all adapter-based methods, the adapter bottleneck dimension is set to 64. All experiments are conducted on a single NVIDIA GeForce RTX 3090 GPU with 24 GB of memory.

\section{Results}\label{results}

We conduct a comprehensive evaluation of PICE through: (1) comparative experiments with SOTA methods on three medical image benchmarks assessing classification performance, stability-plasticity balance, and parameter growth, (2) ablation and mechanistic analyses examining the contributions of individual components, including conditional adapter allocation, classifier recalibration, and task-agnostic adapter fusion, (3) sensitivity analysis to examine model behavior under different inter-task relationships.

\subsection{Comparison with SOTA Methods}\label{comparison-with-sota-methods}

\begin{table*}[!t]
	\caption{Comparison with state-of-the-art continual learning methods on three medical image classification benchmarks. Higher $A_{\mathrm{last}}$ and $A_{\mathrm{avg}}$ and lower $S_{\mathrm{FI}}$ indicate better performance. The best and second-best results are highlighted in bold and underline, respectively. $\dagger$ denotes replay-based methods using stored exemplars.}
	\label{tab:2}
	\centering
	\PICEtablefont
	\setlength{\tabcolsep}{3.0pt}
	\renewcommand{\arraystretch}{1.08}
	\begin{tabular}{llccc|ccc|ccc}
		\toprule
		& & \multicolumn{3}{c|}{Skin8} & \multicolumn{3}{c|}{Kvasir-V2} & \multicolumn{3}{c}{Path16} \\
		Paradigm & Method & $A_{\mathrm{last}}\uparrow$ & $A_{\mathrm{avg}}\uparrow$ & $S_{\mathrm{FI}}\downarrow$ & $A_{\mathrm{last}}\uparrow$ & $A_{\mathrm{avg}}\uparrow$ & $S_{\mathrm{FI}}\downarrow$ & $A_{\mathrm{last}}\uparrow$ & $A_{\mathrm{avg}}\uparrow$ & $S_{\mathrm{FI}}\downarrow$ \\
		\midrule
		Replay-based & CLS-ER$\dagger$~\cite{ref6} & 43.44 & 57.62 & 36.02 & 81.19 & 88.19 & 9.70 & 70.43 & 81.08 & 9.56 \\
		& PCR$\dagger$~\cite{ref7} & 44.40 & 59.97 & 34.68 & 77.12 & 83.83 & 18.57 & 62.24 & 77.06 & 12.44 \\
		Prompt-based & L2P~\cite{ref9} & 33.20 & 56.19 & 38.56 & 34.75 & 62.99 & 37.66 & 18.07 & 44.63 & 59.70 \\
		& DualPrompt~\cite{ref10} & 36.85 & 61.50 & 36.75 & 36.94 & 67.61 & 37.98 & 21.39 & 47.39 & 61.93 \\
		Adapter-based & APER~\cite{ref14} & 49.15 & 64.90 & 33.46 & 84.75 & 89.47 & 11.18 & 72.42 & 81.55 & 7.69 \\
		& EASE~\cite{ref13} & 44.44 & 64.04 & 32.29 & 62.00 & 84.55 & 14.47 & 74.45 & 81.57 & 5.16 \\
		& CRCL~\cite{ref15} & \underline{61.23} & \underline{75.04} & \underline{23.48} & 74.44 & 89.69 & 8.64 & \underline{75.35} & \underline{82.98} & \underline{2.30} \\
		& TUNA~\cite{ref16} & 55.29 & 70.69 & 26.32 & \underline{88.50} & \underline{92.82} & \underline{4.04} & 74.79 & 80.71 & 4.47 \\
		& ACMap~\cite{ref17} & 51.38 & 68.02 & 30.33 & 84.81 & 91.24 & 7.14 & 71.90 & 80.93 & 6.77 \\
		& PICE & \textbf{66.58} & \textbf{77.98} & \textbf{18.53} & \textbf{90.86} & \textbf{94.86} & \textbf{1.74} & \textbf{78.61} & \textbf{84.64} & \textbf{1.11} \\
		\bottomrule
	\end{tabular}
\end{table*}

\subsubsection{\texorpdfstring{\textbf{Overall Classification Performance}}{Overall Classification Performance}}\label{overall-classification-performance}

Table~\ref{tab:2} compares PICE with representative replay-based, prompt-based, and adapter-based CIL methods across three medical image benchmarks. Replay-based methods achieve moderate performance, with CLS-ER reaching 88.19\% \(A_{\text{avg}}\) on Kvasir-V2 but only 57.62\% on Skin8. Although they benefit from direct access to stored exemplars, this reliance on historical images is incompatible with the replay-free setting. Prompt-based methods perform relatively poorly, particularly on Kvasir-V2 and Path16, where the best \(A_{\text{avg}}\ \)reaches only 67.61\% and 47.39\%, suggesting limited adaptation capacity of lightweight prompts for heterogeneous and fine-grained medical distributions.

In contrast, adapter-based methods generally achieve stronger performance by introducing trainable adaptation capacity into intermediate representations. Among adapter-based methods, different capacity organization strategies lead to performance variations. APER and EASE show relatively limited performance under single-adapter reuse and per-task adapter allocation, achieving \(A_{\text{avg}}\) of 64.90\% and 64.04\% on Skin8. CRCL performs best among the baselines on Skin8 and Path16, reaching \(A_{\text{avg}}\) of 75.04\% and 82.98\%, while TUNA achieves the highest baseline result on Kvasir-V2 with 92.82\%. This variation suggests that a fixed adapter organization is not consistently optimal across incremental settings. PICE instead achieves the highest \(A_{\text{last}}\) and \(A_{\text{avg}}\) on all three benchmarks. It achieves its largest relative gain on Skin8, improving \(A_{\text{avg}}\) over CRCL by 3.92\% under the challenging long-tailed and fine-grained setting. The advantage remains consistent on Kvasir-V2 and Path16, despite the visually similar gastrointestinal findings and the substantial cross-task heterogeneity across tissue sources. These results show the consistent effectiveness of PICE across diverse medical distributions.

\begin{figure}[!t]
	\centering
	\includegraphics[width=\columnwidth]{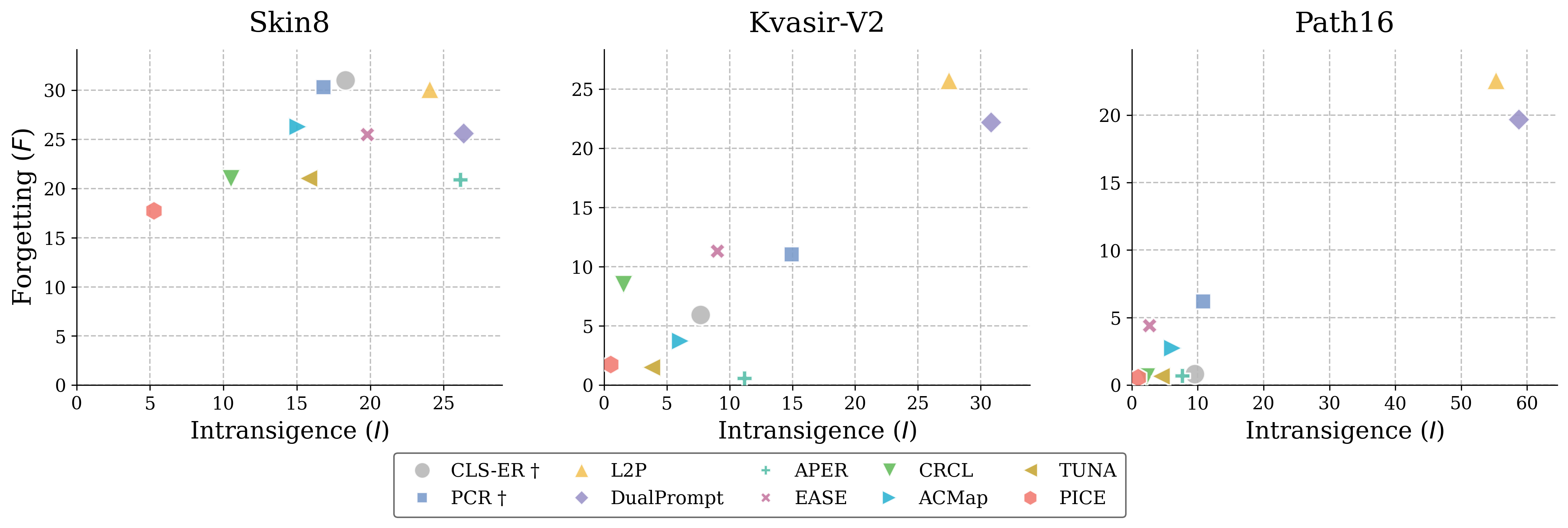}
	\PICEfigurecaptionspacing
	\caption{Forgetting--intransigence comparison on Skin8, Kvasir-V2, and Path16 datasets. The horizontal axis denotes intransigence, which reflects the difficulty of adapting to new classes, while the vertical axis denotes forgetting, which measures degradation on previously learned classes. Points closer to the origin represent a better stability--plasticity trade-off.}
	\label{fig:3}
\end{figure}

\subsubsection{\texorpdfstring{\textbf{Stability--Plasticity Analysis}}{Stability--Plasticity Analysis}}\label{stabilityplasticity-analysis}

To further examine the stability--plasticity behavior behind the aggregated \(S_{\text{FI}}\), Fig.~\ref{fig:3} visualizes the forgetting and intransigence of different CIL methods. On Skin8, the most challenging benchmark, many methods remain relatively distant from the origin, whereas PICE is positioned closest to the lower-left region. This difficulty likely stems from its class imbalance and subtle inter-class differences. On Kvasir-V2, several adapter-based methods approach the origin, with PICE achieving the most favorable overall position, while prompt-based methods remain considerably farther away. The more balanced class distribution may partially explain the reduced forgetting and intransigence observed among stronger methods. Path16 exhibits a distinct pattern, in which most replay-based and adapter-based methods cluster near the origin, whereas prompt-based methods show substantially greater intransigence. The clearer separation among tissue-specific tasks may reduce interference between old and new classes.

\subsubsection{\texorpdfstring{\textbf{Performance across Incremental Stages}}{Performance across Incremental Stages}}\label{performance-across-incremental-stages}

\begin{figure}[!t]
	\centering
	\includegraphics[width=\columnwidth]{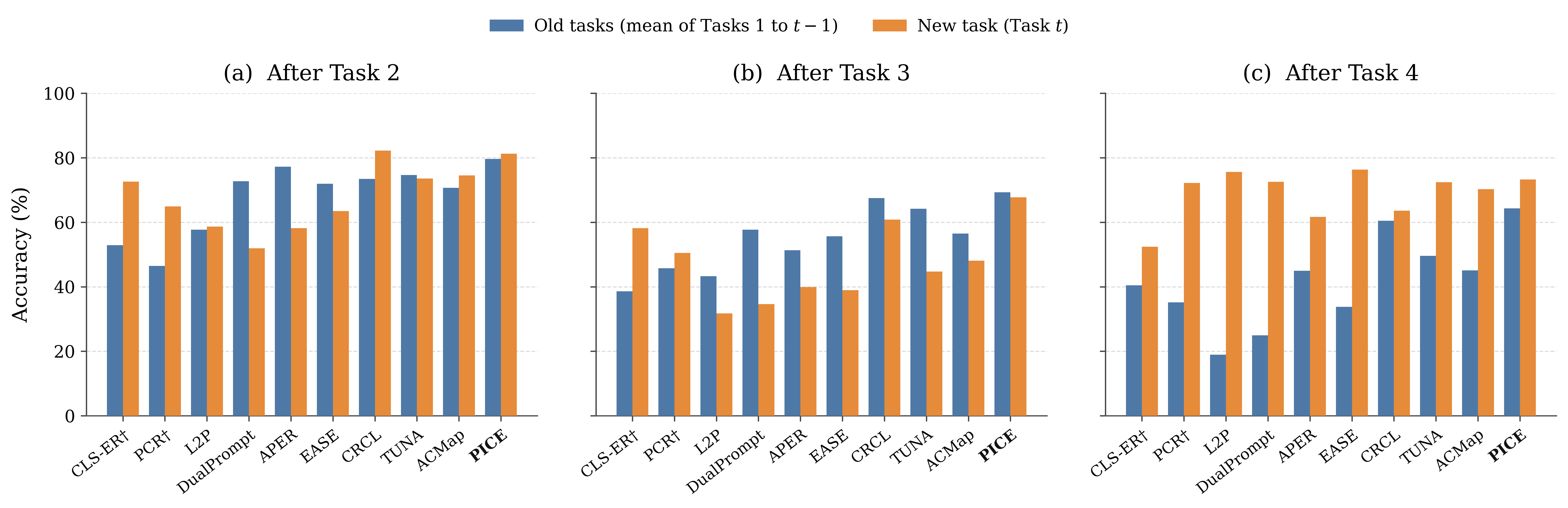}
	\PICEfigurecaptionspacing
	\caption{Old-task retention and new-task adaptation across incremental stages on Skin8. Blue and orange bars denote the average accuracy on old tasks from \(T_{1}\) to \(T_{t - 1}\), and the accuracy on the newly introduced task \(T_{t}\).}
	\label{fig:4}
\end{figure}

Given the more pronounced stability--plasticity challenge observed on Skin8, we select this dataset for a detailed analysis of old task retention and new task adaptation throughout the incremental process. Each panel in Fig.~\ref{fig:4} presents the evaluation after learning \(T_{t}\). The blue bars measure the average accuracy over all previous tasks from \(T_{1}\) to \(T_{t - 1}\), while the orange bars measure the accuracy on the new task \(T_{t}\).

At the beginning of incremental learning, PICE achieves comparable old and new task accuracy of 79.7\% and 81.2\%. This balanced behavior is maintained as more classes are introduced. After \(T_{4}\), EASE and L2P achieve slightly higher current task accuracies of 76.3\% and 75.6\%, but their previous task averages decrease to 33.8\% and 18.9\%. In comparison, PICE retains 64.3\% on previous tasks while achieving a competitive 73.3\% on the current task, maintaining a better balance between knowledge retention and new-class adaptation.

\begin{figure}[!t]
	\centering
	\includegraphics[width=0.6\columnwidth]{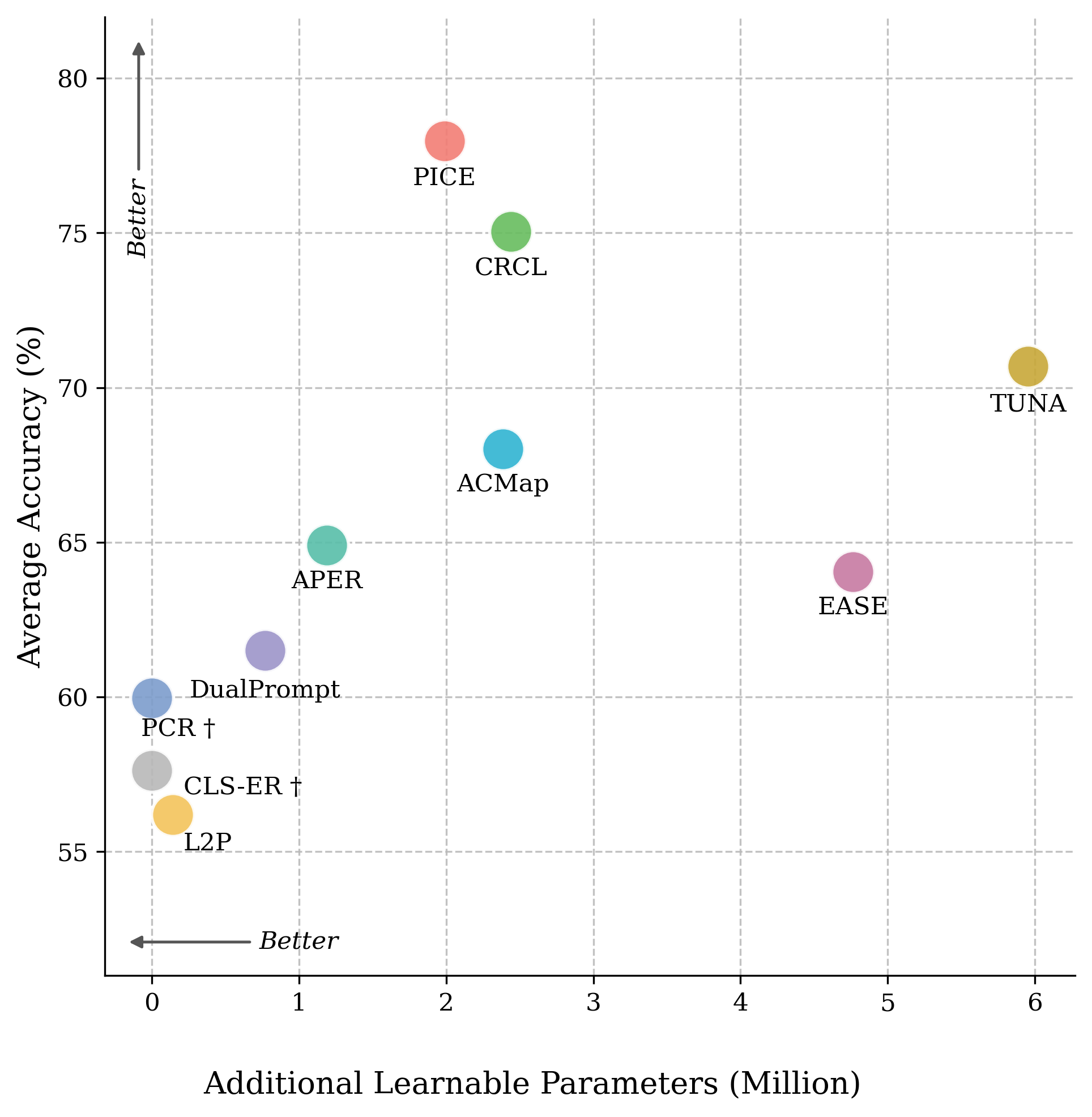}
	\PICEfigurecaptionspacing
	\caption{Accuracy--parameter trade-off of different CIL methods on Skin8. The x-axis denotes the number of additional learnable parameters beyond the frozen pretrained backbone, and the y-axis denotes the average accuracy.}
	\label{fig:5}
\end{figure}

\subsubsection{\texorpdfstring{\textbf{Parameter Growth}}{Parameter Growth}}\label{result-parameter-growth}

Fig.~\ref{fig:5} further compares \(A_{\text{avg}}\) and additional learnable parameters of different CIL methods. Methods located closer to the upper-left region achieve higher accuracy with fewer additional parameters. Replay-based and prompt-based methods introduce few additional parameters but remain below 61.50\% \(A_{\text{avg}}\). Among adapter-based methods, APER limits parameter growth to 1.19 million by reusing a single adapter, yet achieves only 64.90\% \(A_{\text{avg}}\), suggesting that low parameter growth alone does not ensure sufficient adaptation to successive medical distributions. Conversely, allocating more parameters does not necessarily produce better performance. TUNA increases the additional parameter count to 5.95 million, but obtains only 70.69\% in \(A_{\text{avg}}\). CRCL achieves the strongest baseline trade-off, with 75.04\% \(A_{\text{avg}}\) and 2.44 million additional parameters. PICE increases \(A_{\text{avg}}\) to 77.98\% while reducing the parameter count to 1.99 million, corresponding to a 3.92\% \(A_{\text{avg}}\) improvement and an 18.51\% parameter reduction. These results show that PICE achieves stronger performance with limited parameter growth through conditional adapter reuse and expansion.

\subsection{Ablation and Mechanistic Validation}\label{ablation-and-mechanistic-validation}

Skin8 is used for the following analyses because it exhibits the most pronounced stability--plasticity conflict among the three benchmarks. Table~\ref{tab:3} reports a cumulative ablation starting from a frozen ViT baseline and progressively introducing GCR, single-adapter adaptation, multiple accumulated adapters, and prototype-guided fusion. For the Multi-Adapter configuration, adapter fusion is learned by a trainable router, whereas Prototype Fusion replaces the router with prototype-based weighting for task-agnostic inference. The final three rows further compare prototype, linear, and cosine classifier heads under the complete framework.

The cumulative results show consistent improvements as the individual components are progressively incorporated. GCR increases \(A_{\text{avg}}\) from 60.17\% to 67.36\%, reflecting the benefit of recalibrating the classifier with preserved statistical memory. Single-adapter adaptation further raises \(A_{\text{avg}}\) to 73.73\%, showing the benefit of adapting intermediate representations to incoming distributions. Multi-Adapter and Prototype Fusion provide additional gains, with the complete PICE reaching 77.98\% \(A_{\text{avg}}\ \)and reducing \(S_{\text{FI}}\) from 38.01 to 18.53. The classifier-head comparison also shows that the cosine classifier achieves the best overall results among the evaluated classifier heads. Together, these results establish the contribution of the principal components in the complete configuration.

\begin{table}[!h]
	\caption{Ablation study on Skin8. The upper block reports cumulative component ablations, while the lower block compares different classifier heads within the full framework.}
	\label{tab:3}
	\centering
	\PICEtablefont
	\setlength{\tabcolsep}{3.2pt}
	\renewcommand{\arraystretch}{1.08}
	\begin{tabular}{lccc}
		\toprule
		Configuration & $A_{\mathrm{last}}\uparrow$ & $A_{\mathrm{avg}}\uparrow$ & $S_{\mathrm{FI}}\downarrow$ \\
		\midrule
		Baseline & 40.71 & 60.17 & 38.01 \\
		+ GCR & 50.34 & 67.36 & 33.01 \\
		+ Single-Adapter & 60.49 & 73.73 & 21.74 \\
		+ Multi-Adapter & 64.10 & 75.57 & 20.88 \\
		+ Prototype Fusion & \textbf{66.58} & \textbf{77.98} & \textbf{18.53} \\
		\midrule
		Prototype classifier & 44.51 & 61.22 & 31.73 \\
		Linear classifier & 53.87 & 71.60 & 28.09 \\
		Cosine classifier & \textbf{66.58} & \textbf{77.98} & \textbf{18.53} \\
		\bottomrule
	\end{tabular}
\end{table}

\subsubsection{\texorpdfstring{\textbf{Analysis of Conditional Adapter Allocation}}{Analysis of Conditional Adapter Allocation}}\label{analysis-of-conditional-adapter-allocation}

To further examine conditional adapter allocation, we analyze layer-wise responses and the effect of the expansion threshold. Specifically, the former analyzes why evolution is confined to deep adapters, and the latter evaluates how C2ST guides adapter reuse or expansion according to task compatibility.

\begin{figure}[!t]
	\centering
	\includegraphics[width=\columnwidth]{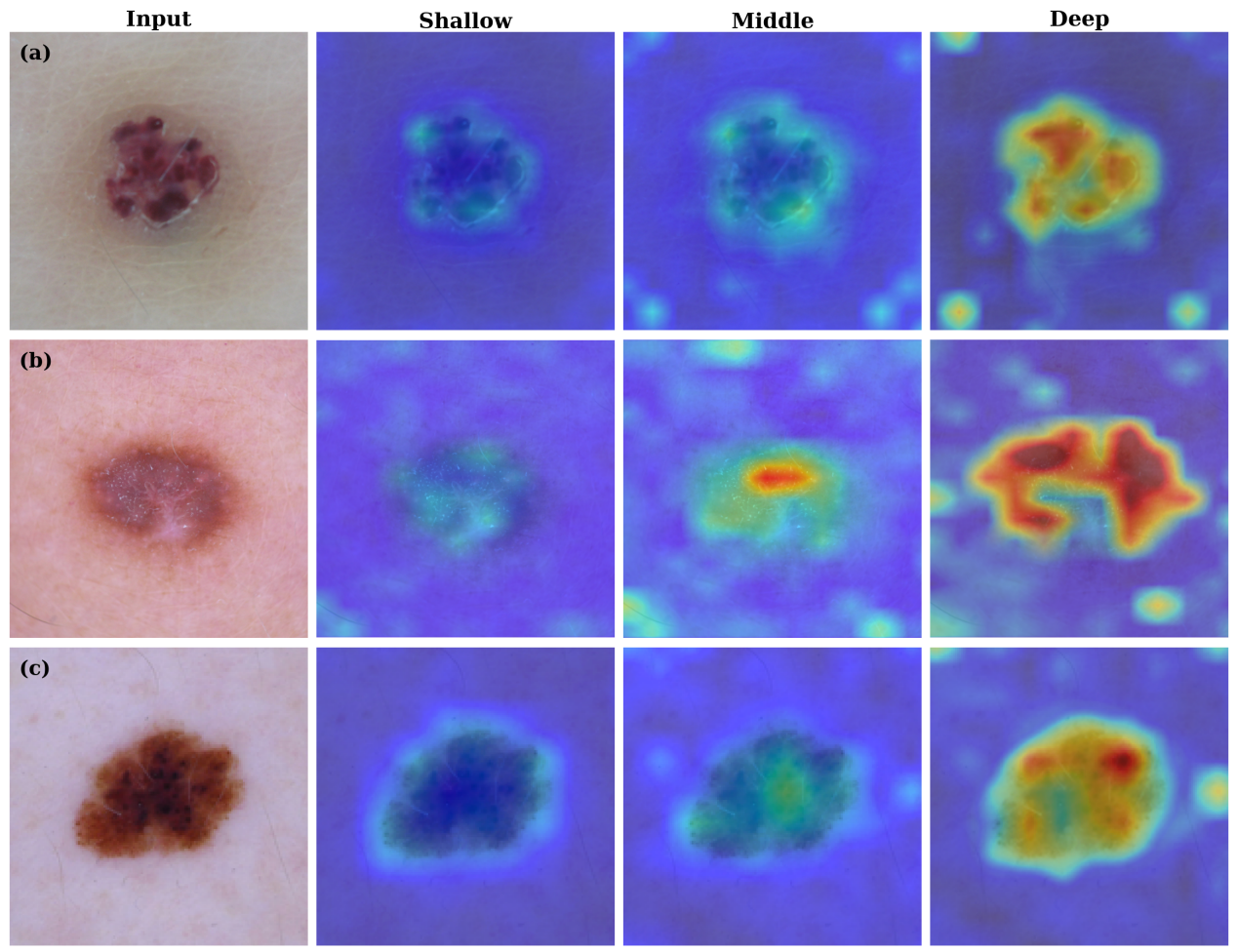}
	\PICEfigurecaptionspacing
	\caption{Layer-wise response visualization of representative Skin8 samples. Each row shows the input image and response maps from shallow, middle, and deep layers.}
	\label{fig:6}
\end{figure}

\textbf{Layer-wise Response Visualization.} To provide a clearer motivation for the layer-aware adapter design, Fig.~\ref{fig:6} visualizes how the attention regions of adapter augmented model change across different network depths. For representative samples, shallow and middle layers respond primarily to broader lesion regions, whereas deeper layers focus more selectively on localized discriminative areas. This layer-wise distinction motivates the architectural design of PICE, with shallow adapters shared and deep expert adapters conditionally reused or expanded.

\begin{figure}[!h]
\centering
\includegraphics[width=\columnwidth]{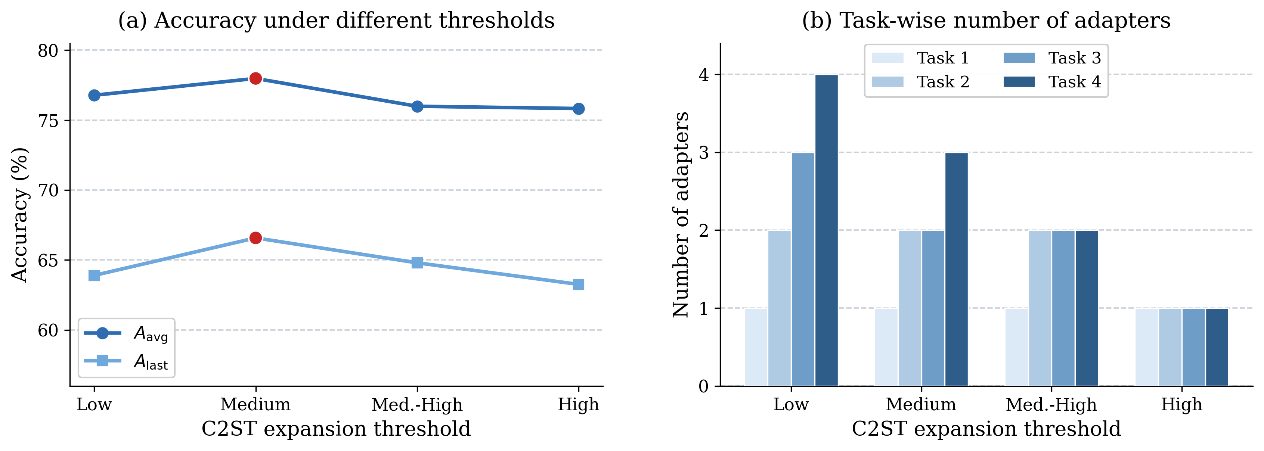}
\PICEfigurecaptionspacing
\caption{Effect of the C2ST expansion threshold on adapter growth. The left panel reports the performance under different threshold settings, and the right panel shows the number of adapters after each incremental task.}
\label{fig:7}
\end{figure}

\textbf{Effect of Expansion Threshold.} To further investigate the behavior of adapter growth, we examine the impact of the predefined discrepancy threshold \(\tau\) in Eq.~\eqref{eq:6}, which controls whether the most compatible old adapter is reused or a new adapter is allocated according to the discrepancy score \(\Delta_{t,k^{\star}}\). As shown in Fig.~\ref{fig:7}, a low threshold triggers expansion at every incremental stage, yielding four adapters but only 76.78\% \(A_{\text{avg}}\). This suggests that assigning a new adapter to each incoming task does not necessarily improve classification performance, as it may limit knowledge sharing between compatible tasks. Conversely, a high threshold suppresses expansion entirely, retaining a single adapter and reducing \(A_{\text{avg}}\) to 75.83\%, indicating that fixed shared capacity is insufficient when the incoming distribution deviates substantially from previously observed ones. The medium-high setting allows limited expansion and achieves 75.99\%. With the medium threshold, PICE expands at \(T_{2}\) and \(T_{4}\) while reusing existing capacity at \(T_{3}\), achieving the best \(A_{\text{avg}}\ \)of 77.98\% with three adapters. These findings indicate that expansion decisions should be guided by compatibility between the incoming distribution and existing adapters, rather than following a fixed expansion rule for every new task.

\subsubsection{\texorpdfstring{\textbf{Analysis of Classifier Recalibration}}{Analysis of Classifier Recalibration}}\label{analysis-of-classifier-recalibration}

Following conditional adapter allocation, newly acquired knowledge must be integrated with previously learned classes. We therefore evaluate classifier recalibration by comparing classifier heads and measuring old--new class bias before and after GCR.

\textbf{Classifier Selection.} To support classification over all observed classes, we compare prototype, linear, and cosine classifiers while keeping all other components unchanged. As shown in Table~\ref{tab:3}, the prototype classifier achieves 61.22\% \(A_{\text{avg}}\), which increases to 71.60\% with the linear classifier and 77.98\% with the cosine classifier. The cosine classifier achieves the best overall performance among the three alternatives. Its normalized decision function may help maintain more consistent classification across incremental stages, and it is therefore adopted in PICE.

\begin{figure}[!h]
\centering
\includegraphics[width=\columnwidth,trim=0 34bp 0 0,clip]{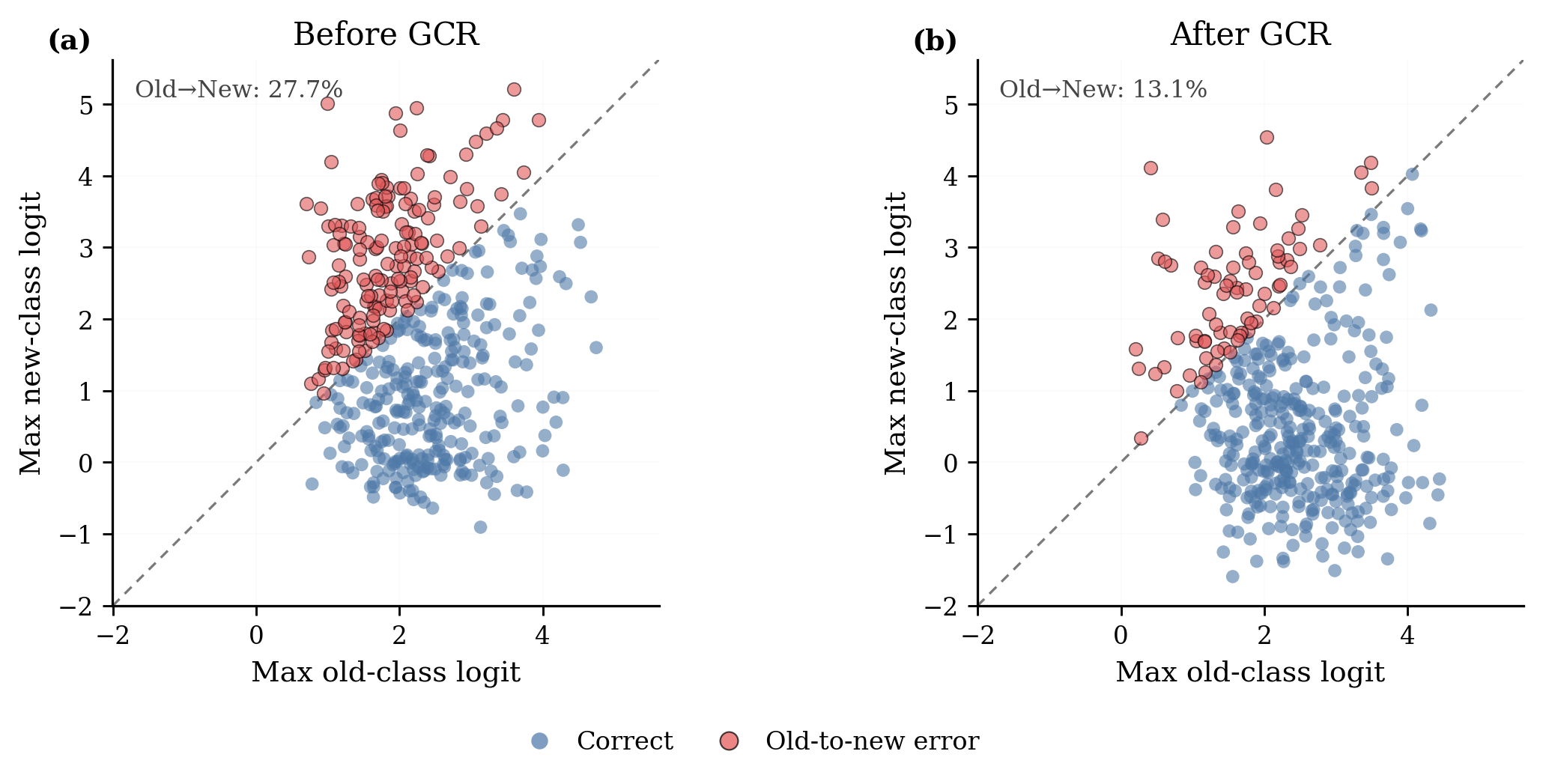}
\PICEfigurecaptionspacing
\caption{Response space visualization before and after GCR. Red points indicate old task samples misclassified as new classes, while blue points denote correct predictions. The dashed line indicates equal maximum logits over old and new classes.}
\label{fig:8}
\end{figure}

\textbf{GCR for Bias Mitigation.} Although the cosine classifier provides a unified decision space, sequential adaptation can still bias predictions toward newly introduced classes. Fig.~\ref{fig:8} compares classifier responses for old task samples before and after GCR. For each sample, the horizontal coordinate is the maximum logit among old classes, denoted as \(s_{\text{old}}\), while the vertical coordinate is the maximum logit among new classes, denoted as \(s_{\text{new}}\). The dashed diagonal \(s_{\text{old}} = s_{\text{new}}\) separates samples with stronger responses to old classes from those with stronger responses to new classes. Before GCR, 27.7\% of old task samples are misclassified as new classes. After GCR, this error rate decreases to 13.1\%, with more old task samples aligned with the old class response region. This result suggests that GCR uses reconstructed features to recalibrate the unified classifier, helping reduce its tendency to favor new classes.

\subsubsection{\texorpdfstring{\textbf{Analysis of Task-Agnostic Adapter Fusion}}{Analysis of Task-Agnostic Adapter Fusion}}\label{analysis-of-task-agnostic-adapter-fusion}

Following conditional evolution, we further examine how the accumulated adapters are utilized without task identity. Specifically, we analyze prototype-guided adapter usage and the effect of the Top-\(K\) fusion strategy.

\begin{figure}[!h]
\centering
\includegraphics[width=0.68\columnwidth,trim=0 5bp 0 0,clip]{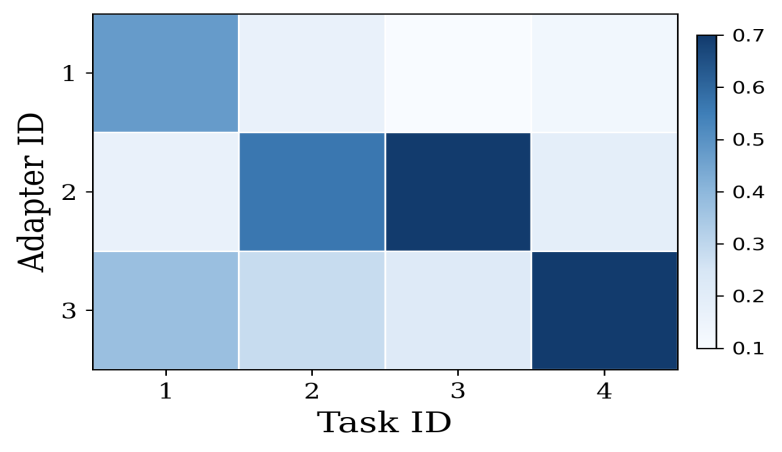}
\PICEfigurecaptionspacing
\caption{Adapter usage across incremental tasks on Skin8. Darker colors indicate higher prototype-guided routing weights.}
\label{fig:9}
\end{figure}

\textbf{Adapter Usage.} Fig.~\ref{fig:9} visualizes the adapter usage of each task on the test set. A clear structured routing pattern emerges, in which \(T_{1}\) mainly relies on Adapter 1, \(T_{2}\) and \(T_{3}\) prefer Adapter 2, and \(T_{4}\) is primarily routed to Adapter 3. This pattern is consistent with the adapter allocation decisions made during training, which supports the use of prototype similarity to recover the learned capacity structure without task identity. The remaining nonzero weights reflect soft routing, allowing relevant adapters to contribute when appropriate.

\begin{table}[!h]
\caption{Ablation of the Top-$K$ fusion strategy on Skin8, where $K$ denotes the number of selected task-associated adapter collections. Different tasks may share the same adapter through conditional reuse.}
\label{tab:4}
\centering
\PICEtablefont
\setlength{\tabcolsep}{1.5pt}
\renewcommand{\arraystretch}{1.08}
\begin{tabular}{lccccc}
\toprule
Fusion Strategy & $A_{\mathrm{last}}\uparrow$ & $A_{\mathrm{avg}}\uparrow$ & $F\downarrow$ & $I\downarrow$ & $S_{\mathrm{FI}}\downarrow$ \\
\midrule
Top-1 adapter & 63.87 & 76.20 & 20.87 & 5.79 & 21.66 \\
Top-2 adapters & 65.95 & 77.01 & 18.18 & 7.20 & 19.55 \\
Top-3 adapters & \textbf{66.58} & \textbf{77.98} & 17.76 & \textbf{5.27} & 18.53 \\
Top-4 adapters & 65.52 & 77.54 & \textbf{17.23} & 6.76 & \textbf{18.51} \\
\bottomrule
\end{tabular}
\end{table}

\textbf{Top-}\(\mathbf{K}\) \textbf{Fusion Strategy.} We vary \(K \in \{ 1,2,3,4\}\) to evaluate routing sparsity during task-agnostic inference, treating each task as a separate candidate even when expert adapters are shared. As shown in Table~\ref{tab:4}, Top-1 achieves the lowest \(A_{\text{avg}}\) of 76.20\%, and the highest forgetting of 20.87. This indicates that hard adapter selection may be insufficient, since a test sample may share visual similarities with multiple previously learned tasks. Expanding fusion scope from Top-1 to 3, the accuracy improves steadily, as several relevant adapters can jointly exploit reusable knowledge across tasks. Although Top-4 slightly reduces forgetting, it leads to lower \(A_{\text{avg}}\), suggesting that including additional adapters may dilute the contribution of the most informative ones.

\subsection{Sensitivity to Different Inter-Task Relationships}\label{sensitivity-to-different-inter-task-relationships}

\begin{figure}[!h]
	\centering
	\includegraphics[width=\columnwidth]{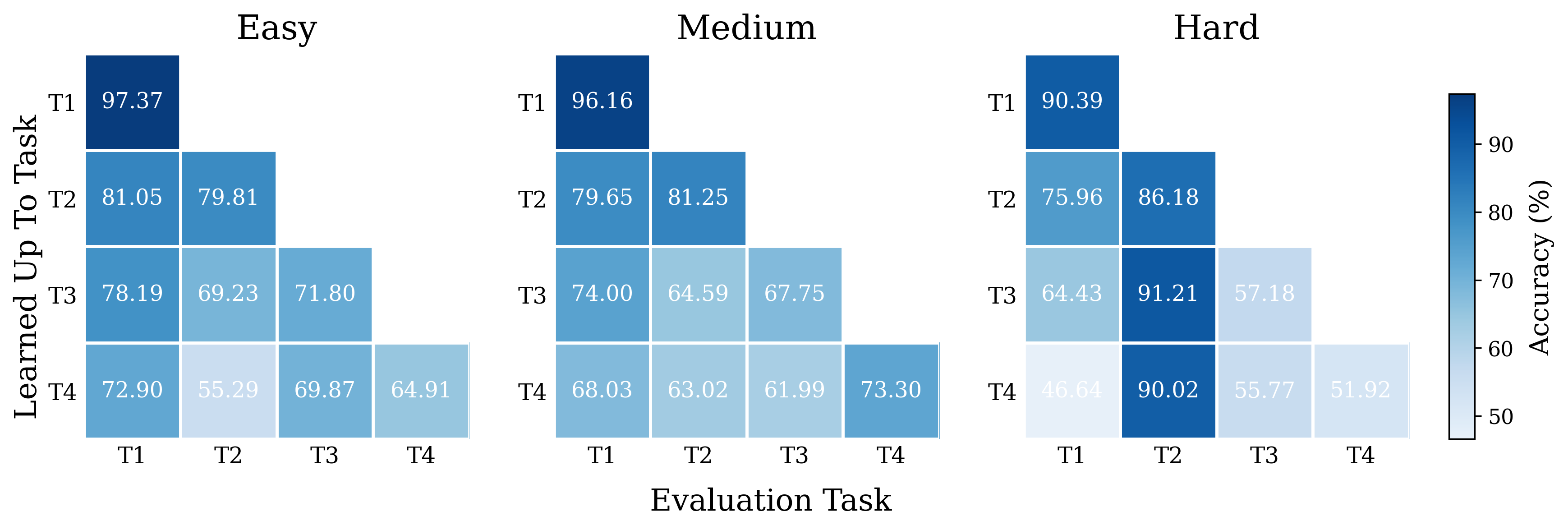}
	\PICEfigurecaptionspacing
	\caption{Accuracy matrices under different task difficulty levels on Skin8. Rows indicate the learning stage and columns indicate the evaluated task.}
	\label{fig:10}
\end{figure}

To examine whether our method is robust to different task difficulties, we conduct an additional evaluation on Skin8 following the main idea of the EDGE protocol~\cite{ref30}. In the Easy setting, adjacent tasks are more similar, reducing distributional gaps and facilitating knowledge reuse. In contrast, the Hard setting places more dissimilar tasks next to each other, making knowledge transfer more difficult and increasing forgetting. As shown in Fig.~\ref{fig:10}, our method remains relatively stable under the \emph{Easy} and \emph{Medium} settings. Under the more challenging \emph{Hard} setting, the final retained accuracy \(A_{\text{last}}\) decreases to 61.09\%, while the overall average accuracy remains 75.87\%. Performance degrades moderately as task difficulty increases, while PICE continues to retain prior knowledge and adapt to new tasks across different task orders.

\section{Discussion}\label{discussion}

To support self-evolving medical image recognition under privacy and deployment constraints, we propose PICE, a prototype-informed conditional model evolution framework as new disease categories emerge. Experiments across three medical benchmarks demonstrate that PICE consistently improves incremental classification performance and the stability--plasticity balance while limiting parameter growth. We next discuss what knowledge should be preserved, when additional model capacity is necessary, and how newly acquired knowledge can be consolidated and effectively reused.

\subsection{What to Preserve: Toward Knowledge Preservation without Historical Images}\label{what-to-preserve-towards-knowledge-preservation}

Preserving historical knowledge without retaining patient images is a critical requirement for the clinical application of self-evolving medical AI systems. Replay-based methods such as CLS-ER and PCR preserve prior knowledge through stored exemplars, but their reliance on historical patient data is less suitable under privacy and data-sharing constraints. PICE instead uses prototype-statistical memory to represent each learned disease through its prototype and covariance to capture the distribution center and intra-class variation. As shown in Table~\ref{tab:2}, PICE achieves an \(A_{\text{avg}}\) of 77.98\% on Skin8 without retaining historical images, substantially exceeding the 57.62\% and 59.97\% obtained by CLS-ER and PCR. 

Beyond preserving prior knowledge, this unified design also ensures that the knowledge used to determine model evolution remains consistent with that used for consolidation and inference. Prototype-statistical memory thus acts as a common interface between preserved disease knowledge and evolving model capacity, allowing historical knowledge to influence not only what the model remembers, but also how it grows and how its accumulated capabilities are used.

\subsection{When to Evolve: Toward Conditional Capacity Allocation}\label{when-to-evolve-towards-conditional-capacity-allocation}

A central challenge in self-evolving medical image recognition is aligning model capacity growth with the representational demands of emerging disease distributions. However, existing approaches typically apply the same strategy to every new task, regardless of its relationship to previously learned distributions. Using a single shared adapter may force heterogeneous disease distributions into the same representation space, whereas assigning a new adapter to each task overlooks transferable representations and leads to redundant parameter growth. As shown in Table~\ref{tab:2} and Fig.~\ref{fig:5}, APER achieves 64.90\% \(A_{\text{avg}}\) with 1.19M additional parameters under single adapter reuse, whereas EASE introduces 4.76M parameters through task-specific expansion but reaches an  \(A_{\text{avg}}\) of 64.04\%.

In contrast, PICE treats adapter allocation as a distribution-compatibility problem, using C2ST to determine whether existing capacity can be reused or additional capacity is required. This is reflected in the Skin8 experiments, where \(T_{2}\) and \(T_{4}\) trigger expansion while \(T_{3}\) reuses existing adapter. The threshold analysis in Fig.~\ref{fig:7} further shows that forcing frequent expansion and excessive reuse both reduce classification performance, confirming that neither maximum specialization nor maximum sharing is optimal. PICE also confines structural evolution to deeper Transformer blocks while sharing shallow adaptation capacity across tasks. As illustrated in Fig.~\ref{fig:6}, shallow representations respond to broader lesion regions, whereas deeper layers increasingly focus on fine-grained discriminative areas, supporting the use of shared shallow adapters and selectively expandable deep adapters. This selective evolution supports clinically maintainable self-evolving medical image recognition by accommodating emerging disease distributions without unnecessary model growth.

\subsection{How to Evolve: Toward Knowledge Consolidation and Effective Reuse}\label{how-to-evolve-towards-knowledge-consolidation-and-effective-reuse}

Structural adaptation alone, however, is insufficient for continual model evolution. Existing task-specific adapter strategies can preserve specialized representations across tasks, but they do not directly address the bias between old and new classes in a unified decision space. PICE therefore realizes model evolution through task adaptation followed by memory updating and classifier recalibration, so that newly acquired knowledge can be integrated with previous classes. As shown in Fig.~\ref{fig:8}, GCR reduces the proportion of old-task samples misclassified as new classes from 27.7\% to 13.1\%, demonstrating a substantial reduction in old--new class bias. This behavior suggests that the value of consolidation lies not in adding further capacity, but in integrating new diagnostic capabilities into a coherent model across successive learning stages.

Effective evolution also requires accumulated capacity to remain accessible when task identity is unavailable. Simply retaining multiple adapters may fragment knowledge across task-specific modules, while hard selection can overlook useful information shared across related tasks. PICE uses preserved knowledge to identify and combine relevant adapters during task-agnostic inference. According to Table~\ref{tab:4}, Top-3 fusion improves \(A_{\text{avg}}\) from 76.20\% to 77.98\% over Top-1 selection. This is further supported by the routing patterns in Fig.~\ref{fig:9}, where related tasks tend to rely on the same dominant adapter, consistent with the allocation decisions made during training. These results indicate that accumulated diagnostic capacity remains recoverable and reusable during inference.

\subsection{Clinical Implications, Limitations, and Future Work}\label{clinical-implications-limitations-and-future-work}

Beyond improving incremental recognition performance, PICE suggests a shift from one-time model deployment toward self-evolving medical AI. By decoupling knowledge retention from access to historical patient images, it offers a way to preserve diagnostic continuity under privacy constraints. Its conditional model evolution further changes model updating from routine parameter expansion to selective reorganization of existing and newly acquired capacity, which is important for controlling deployment complexity as diagnostic taxonomies evolve. Its consistent performance across three imaging modalities further suggests the potential for extensible and clinically maintainable medical AI across different medical settings.

A remaining limitation is that PICE assumes emerging disease categories are predefined before incremental learning and therefore cannot autonomously identify previously unseen categories from incoming clinical data. In real world deployment, however, self-evolving medical AI should ideally recognize when newly observed patterns cannot be adequately explained by existing disease knowledge and determine whether they represent genuinely emerging categories that require model evolution. Future work will investigate automatic discovery and verification of unknown disease categories, enabling the model to trigger subsequent knowledge acquisition and capacity evolution with less reliance on predefined task boundaries.

\section{Conclusion}\label{conclusion}

In this work, we proposed PICE, a replay-free framework that enables medical image classifiers to evolve when existing capacity is insufficient. By using prototype-statistical memory to connect conditional adapter allocation, knowledge consolidation, and task-agnostic inference, PICE integrates structural evolution and knowledge preservation within a unified design. Experiments on Skin8, Kvasir-V2, and Path16 show consistent improvements in incremental accuracy and stability-plasticity balance while limiting parameter growth. These results support PICE as a practical step toward updatable and clinically maintainable medical AI under privacy constraints.

\end{document}